\documentclass[intlimits,twoside,a4paper]{article}

\usepackage[cp1251]{inputenc}
\usepackage{xcolor}
\usepackage[eqsecnum]{cmpj3}

\usepackage{bm}

\newcommand{\bea}{\begin{eqnarray}}
\newcommand{\eea}{\end{eqnarray}}
\newcommand{\be}{\begin{equation}}
\newcommand{\ee}{\end{equation}}

\issue{2021}{24}{1}{13602}
\doinumber{10.5488/CMP.24.13602}

\title[Correlation between kinetic fragility and Poisson's ratio from
analysis of data for soft colloids]{Correlation between kinetic fragility and Poisson's ratio from
analysis of data for soft colloids}

\author[A. Mondal, L. Premkumar,  S.P. Das]{A. Mondal\refaddr{label1},
L. Premkumar\refaddr{label1, label2},  S.P. Das\refaddr{label1}
\footnote{corresponding author, email address: shankar0359@yahoo.com}} 

\addresses{\addr{label1} School of Physical Sciences, Jawaharlal
Nehru University, New Delhi 110067, India
\addr{label2} Department of Physics, National Institute of Technology
Manipur, Imphal 795004, India}

\date{Received August 10, 2020, in final form November  10, 2020}
\begin{document}

\maketitle
\begin{abstract}
We consider the link between fragility and elasticity that follows
from the analysis of the data for a set of soft colloid materials
consisting of deformable spheres reported by [Mattsson { et
al.}, Nature, 2009, {\textbf 462}, 83]. 
The present work makes a
quantitative analysis through an explicit
definition for fragility index $m$  in terms of density
dependence, extending the corresponding formula  of $m$ for molecular
systems in terms of temperature dependence. 
In addition,
we fit the data for the high-frequency shear modulus for the
respective soft-colloid to a corresponding theoretical expression
for the same modulus. This expression for the elastic constant is
in terms of the corresponding pair correlation function for the
liquid treated as of uniform density. The pair correlation
function is adjusted through a proper choice of the parameters for
the two body interaction potential for the respective soft-colloid
material. The nature of correlation between the fragility and
Poisson ratio observed for the soft colloids is  qualitatively
different, as compared to the same for molecular glasses. The
observed link between fragility of a metastable liquid and its
elastic coefficients is  a manifestation of the effects of
structure of the fluid on its dynamics.
The present work thus analyses the data on soft colloids and by introducing
definitions from statistical mechanics obtains a correlation between kinetic fragility and
Poissons's ratio for the soft material.

\end{abstract}



\section{Introduction}

An instructive plot of the final relaxation time $\tau_\alpha$ (on
a logarithmic scale) of supercooled liquids was made by Angell
\cite{angell,turnbull} with respect to the inverse temperature
scaled with $T_\mathrm{g}$,  i.e., with $x=T_\mathrm{g}/T$.
For any specific system, the temperature $T_\mathrm{g}$  is
defined to be the one at which the relaxation time $\tau_\alpha$
grows by a chosen order of magnitudes ${\cal B}$ (say) over a
characteristic short time value. The quantity ${\cal B}$ is the same
for all materials and generally it is chosen to be $16$
\cite{bohmer1, bohmer2} in molecular systems. Therefore, the Angell
plot, by construction, has all the relaxation curves merging at a
single point on the $y$ axis at $y={\cal B}$ and at $x=1$, 
i.e., for $T=T_\mathrm{g}$. Slope of the relaxation time
$\tau_\alpha$ (on a log scale)  vs scaled temperature $x$ at
$T=T_\mathrm{g}$ is termed as the fragility index $m$.
For fragile liquids, $\tau_\alpha$ are very sensitive to
temperature changes near  $T_\mathrm{g}$, and $m$ is large. Si${\rm
O}_2$ and $o$-terphenyl respectively denote two extreme cases of
strong and fragile systems with $m$ values 20 and 81.

Soft matter \cite{rev-sid,rev-ajay,rev-bonn,gnan} has constituent
elements of much larger size than that of the so-called ``hard
matter'' consisting of atomic size particles. The soft matter is
characterized by small values for its elastic moduli. The nature of the glassy
dynamics for molecular systems as depicted above is also observed
in the relaxation behaviour of soft matter
\cite{sidnagel-prl,nagel-epl,weeks-review,weitz-review}.
The concentration dependence of relaxation time and elastic
properties of deformable colloidal particles have been analyzed at a
fixed temperature.  The soft colloid materials studied were
aqueous suspensions of deformable microgel particles of varying
elasticity. The microgel particles consisted of interpenetrated
and cross-linked polymer networks of poly (N-isopropylacrylamide)
and polyacrylic acid.  A key thermodynamic property controlling
both the elastic and relaxation properties of the soft
colloids is the concentration variable $\zeta$. The density of
colloid particles, denoted as $\rho$, is determined in terms of
polymer concentrations of the microgel suspension. For a system of
hard spheres of diameter $d$, the particle concentration $\zeta$
is the same as the packing fraction~$\phi$. However, since the
microgel particles considered here are deformable, the packing
fraction $\phi$ is not simply proportional to the number density. For
such systems, we work with the concentration defined as
$\zeta={\rho}v_0$, with $v_0$ being the volume of an undeformed
sphere. In the hard-sphere limit $v_0=\piup{d^3}/6$, making
$\phi\equiv\zeta$.  In the model for deformable soft colloids that
we present here, physical quantities like relaxation times or
shear modulus are plotted as functions of relative concentrations
$(\zeta/\zeta_\mathrm{g})$ or $(\zeta/\zeta^*)$, respectively in
terms of concentration  at the so-called glass transition point
($\zeta_\mathrm{g}$) and at a crossover point ($\zeta^*$).  Thus,
the role of $v_0$ drops out. For simplicity, we are considering a
mono-dispersed system, hence size ratio of the particles does not
enter, though the present approach can be applied to mixtures as
well.

For three different soft colloid materials, logarithms of the
relaxation time $\tau_\alpha$ were measured using different
methods and plotted \cite{weitz} against relative concentration
$\zeta/\zeta_\mathrm{g}$ scaled with respect to its value at the
so-called glass transition point  $\zeta_\mathrm{g}$. The latter
is defined  for soft colloids, in close analogy with
$T_\mathrm{g}$  for the molecular systems. It is chosen so that
the relaxation time $\tau_\alpha$ compared to a microscopic scale
$\tau_0$ grows by a factor ${\cal B}$:
\be
\label{defnB}
\log_{10}\left
[\frac{\tau_\alpha(\zeta_\mathrm{g})}{\tau_0}\right ]={\cal B}.
\ee
The result is an Angell-like plot with end points of all three
relaxation curves meeting at a single point on the Y-axis, at
$x=1$,  and $y{\equiv}{\cal B}=5$.
Apart from the specific value of ${\cal B}$, primarily arising
from technical  limitations of measuring relaxation times with
light scattering, the relative variations  within the group of
different soft-matters closely resemble  those of molecular
systems.  At a qualitative level, soft matters exhibit similar variations in fragility as
that for molecular liquids at fixed volume \cite{angell}. The
difference in the value of ${\cal B}$ for the respective groups also
categorizes the glass-forming materials into different classes.
For the relaxation times covered in the present Angell plot for
soft-colloids, the system  is not in a jammed state even at
$\zeta_\mathrm{g}$, and the Brownian force signifying thermal noise is
present.

The fragility $m$ in this case is defined with respect to
the concentration variable:
\be \label{frag-defn}
m=\frac{d\log_{10}{\tau_\alpha}}{d(\zeta/\zeta_\mathrm{g})}{\Big
    |}_{\zeta=\zeta_\mathrm{g}}. \ee
While the relaxation time signifies the liquid-like behaviour,
elasticity is a characteristic property related to the rigidity of
a system with localized particles. In a disordered fluid state, the
elastic behaviour only persists for short times after a stress is
applied. The corresponding elastic constant is denoted as
$G_\mathrm{p}(\zeta)$~\cite{osuji}. For a set of soft colloids, the
plot of $G_\mathrm{p}(\zeta)$ vs scaled concentration
$\zeta/\zeta^*$  is observed to be  similar to that of
high-frequency shear modulus $G_\infty(T)$ vs scaled temperature
$T^*/T$ for a molecular system~\cite{weitz}.
Here, the reference $\zeta^*$ is identified as the concentration
value over which the stretching exponent for the decay function
(of William-Watts form) remains constant. This would
signify time temperature superposition.  The concentration
variable $\zeta$ replaces the temperature in this case. What
matters in the present context is its value relative to the
scaling density $\zeta/\zeta^*$.
Concentration dependence of the elastic constant
$G_\mathrm{p}(\zeta)$ for three different soft colloids
\cite{weitz} is analysed here. From a theoretical perspective
\cite{mz}, high-frequency elastic constants are related to basic
interaction of the particles and in case of pairwise additive
interaction potentials, results for high-frequency elastic
constants are obtained in terms of pair correlation functions
$g(r)$ within simplest approximations. Our main focus here is on
the structural aspects and we model them in terms of simple two-body interaction potentials.

We  focus on two properties of the soft colloids, depicted above,
i.e., the fragility and short time elastic response in the
soft matter.  Relating fragility of a glass-forming system to the
interaction potential~\cite{casalini,mohanty,casalini1} has been a
topic of much interest. Theoretical studies using landscape
paradigm \cite{Wales,pes-frag} of a many particle system have also
been made to further understand this link. Microscopic models for
the glassy state, based on either thermodynamic
\cite{singh-woly,hartmut,kaur-pre} or  dynamic approach
\cite{mpp,bpeak-spd,sudha} simply assume the solid-like properties
for the amorphous state. In the  thermodynamics based formulation,
the metastable states of a disordered solid are described
\cite{trazona,trk}   in terms of localized density profiles which
also signify vibrational modes \cite{trk,nbpk_jcp}. For dynamical
models, solid-like behavior for the non-ergodic amorphous state
\cite{mezler-prl} is manifested through introduction of Goldstone
type modes \cite{mp-spd,mpp,novikov}. Correlations between
fragility and elastic properties of glassy systems is also
observed in the studies involving the non-ergodicity parameter~\cite{w12}, the bulk and shear moduli
\cite{w10,w11,sokolov-nature,novikov2}, and the strength of the
boson peak~\cite{bpeak-spd,bpeak-plas}. Both of the above two
theoretical descriptions of the amorphous state, include the
static structure factor which plays a key role in understanding
the elastic and dynamic properties.


Reference \cite{weitz} made an important observation that the data on
soft colloids produce a plot similar to what is known as
Angell-plot for molecular systems. The present work makes a
quantitative analysis of these data through an explicit
definition for fragility index $m$  in terms of density
dependence, extending the corresponding formula  of $m$ for molecular
systems in terms of temperature dependence. Thus, we are able to
charactaerize each soft colloid material with a value of $m$.
We also identify the optimum parameters for interaction
for the corresponding solft colloid materials by fitting the experimental data  for
shear modulus $G$ with the formula obtained using basic statistical
mechanics. The bulk modulli $B$ is
obtained for the same interaction by evaluating the corresponding
formulae for the same that follows from the model used for $G$. Thus, Poisson's ratio
$\nu$ for each material is also calculated.  This makes it possible for
the present analysis to also study the frequently studied correlation 
between the  fragility $m$ and  Poisson's ratio
$\nu$, and this observed correlation is found to be qualitatively contrary to the corresponding result for  molecular systems.  This  new aspect of the ``Angell Plot'' is
something we obtain from our analysis here.
In our view, this apparent reversal of the trends in correlations
between $\nu$ and $m$ does not present any contradiction and  can also be a consequence of this
new type of plots and the corresponding definition of the
fragility index as explained above.
The paper is organized as follows. In the next section, we discuss
the calculation of fragility index from the relaxation curves for
the soft colloids. In  section III we discuss the theoretical
models for short time elastic coefficients in terms of the static
correlations.  We briefly describe here the input static
correlation functions and how the latter is controlled by
parameters of the soft sphere interaction potentials.  In section
IV we discuss the comparison of elastic coefficient data for 
different colloids and link them to the corresponding fragility
parameters. We end the paper with a discussion of results and
indicate the limitations of our conclusions.

\section{Angel plot: fragility index}

In the present paper the two basic properties for  soft colloid materials
on which  we focus our analysis, are respectively the fragility
index related to the long-time relaxation behaviour and the
Poisson ratio related to the high-frequency or equivalently short-time elastic response of the soft matter. In the present section,
we discuss the calculation of the first quantity, {i.e.}, the
fragility index, using the Angell plot of the relaxation data
reported.   In the next section we consider the model for high
frequency or short-time elastic constants.

\subsection{Angel plot for soft colloids}
\begin{figure}[!b]
    \centering
    \includegraphics[width=0.55\textwidth]{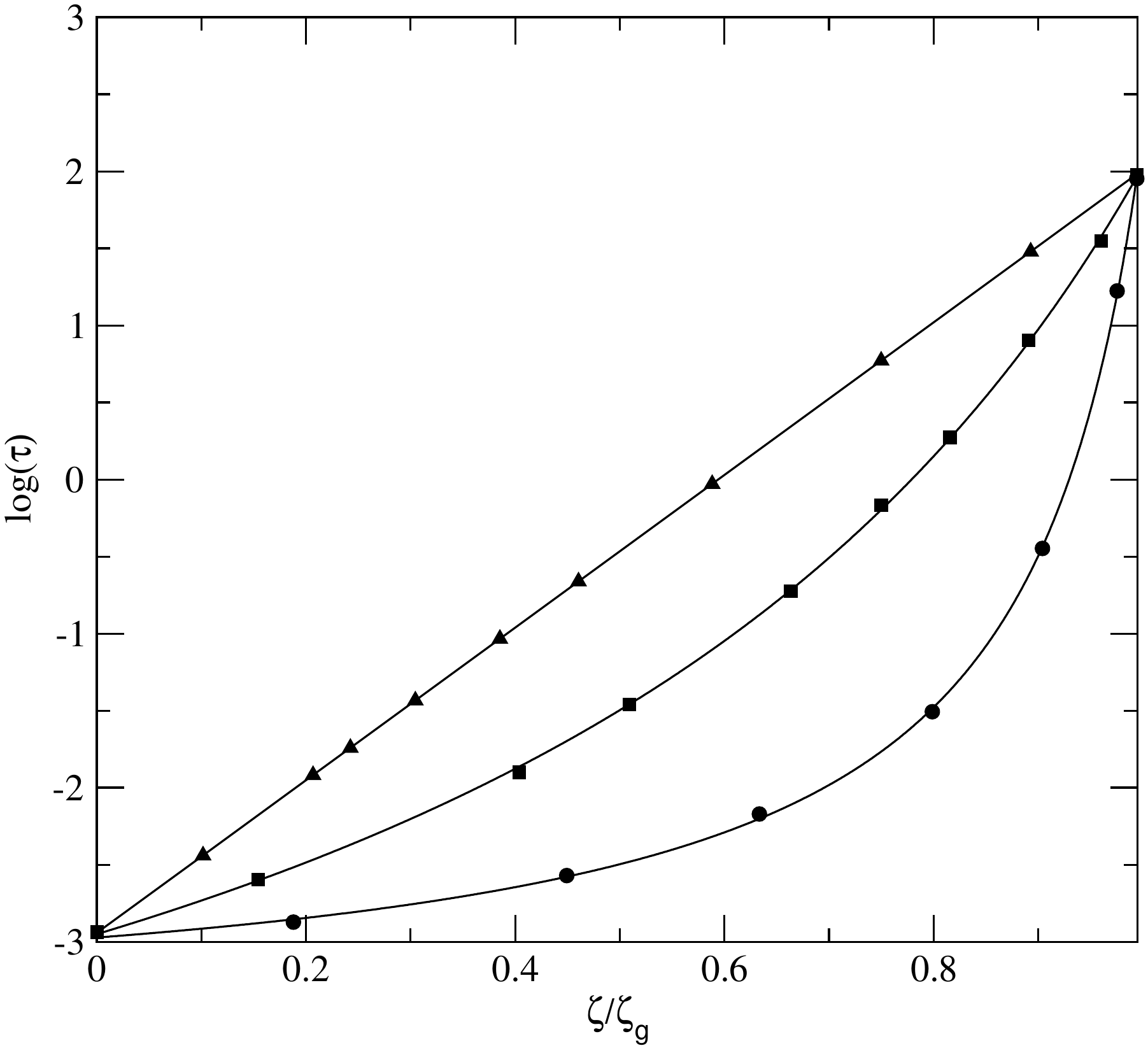}
\caption{The Angell plot with the relaxation data for three soft
colloid materials I, II, and III from  \cite{weitz}, shown with
filled circles, squares, and triangles, respectively. Fits with the
form (\ref{vft-form}) in text for ${\cal B}=5$ and different choice
for $\kappa=\zeta_\mathrm{K}/\zeta_\mathrm{g}$  are shown with
solid lines. Corresponding  $\kappa$ values are listed in table~\ref{table02}, column 6.}
\label{fig01}
\end{figure}

To compute the fragility index for a soft colloid material, we analyze
the density dependence of the relaxation time $\tau_\alpha$ by
plotting (on a log scale) the relaxation time   with respect to
the concentration~$\zeta$, introduced in the previous section.
This so-called Angell  plot \cite{angell,turnbull,apl1,apl2} was shown in figure~1b of~\cite{weitz} for three different soft
colloids, by showing the respective $\log_{10}\left [
{\tau_\alpha}/{\tau_0} \right ]$ vs the relative concentration
$\zeta/\zeta_\mathrm{g}$. For discussions of the present work, we
refer to the three different soft colloid materials  considered here 
 as SC-I, II, and III, respectively. In the Angell plot,
the relaxation curves for three respective materials  meet at a
single point  $\zeta/\zeta_\mathrm{g}\equiv{x}=1$. The so-called
glass transition point $\zeta_\mathrm{g}$ was already defined in
the previous section with the relation (\ref{defnB}) and in the
present case the growth in relaxation time is ${\cal  B}=5$.
Assuming that  each of the respective relaxation curves for the
soft colloids SC-I, II, and III follows a Vogel-Fulcher-like form,
the variation of $\tau$ with $\zeta$ is written as
\be \label{vft-form} \tau_\alpha=\tau_0\exp \left [
\frac{A\zeta}{\zeta_\mathrm{K}-\zeta}\right ]. \ee
The concentration $\zeta_\mathrm{K}$ signifies divergence of the
relaxation time and is often identified with the Kauzmann point for
a glass-forming material.
The fragility index for the soft colloids is defined by a relation
similar to equation~(\ref{frag-defn}) for molecular systems.

With the above definition of the fragility, using the relations
(\ref{defnB}), and (\ref{vft-form}) we obtain the relation:
\be \label{fragil-we} m=\frac{\cal B\kappa}{\kappa-1}.\ee
$\kappa=\zeta_\mathrm{K}/\zeta_\mathrm{g}$ is the ratio of
respective concentrations  $\zeta$ at the so-called Kauzmann point,
signifying the divergence of characteristic relaxation time, to the
value of $\zeta$ at the glass transition point. Equation
(\ref{vft-form}) for the relaxation time can now be presented in
the form:
\be \label{rel-fit} \log_{10} \left [
\frac{\tau_\alpha(x)}{\tau_0}
\right ] = {\cal B}x \frac{\kappa -1}{\kappa -x}.\ee
For a chosen ${\cal B}$ characterizing the Angell plot, the
parameter $\kappa$  attains a specific value corresponding to each
relaxation curve, to obtain the $\tau_\alpha(x)$ vs $x$ behaviour
for the respective soft-colloid materials in the Angell plot. Using $\kappa$
in the relation (\ref{fragil-we}) we obtain the fragility of the
corresponding material.  The fragility also depends on  the chosen
value for the constant ${\cal B}$. For soft colloids this number
is generally smaller than that for a molecular liquid.

\subsection{Fragility index for the soft colloids}
\begin{figure}[!t]
\centering
\includegraphics[width=0.5\textwidth]{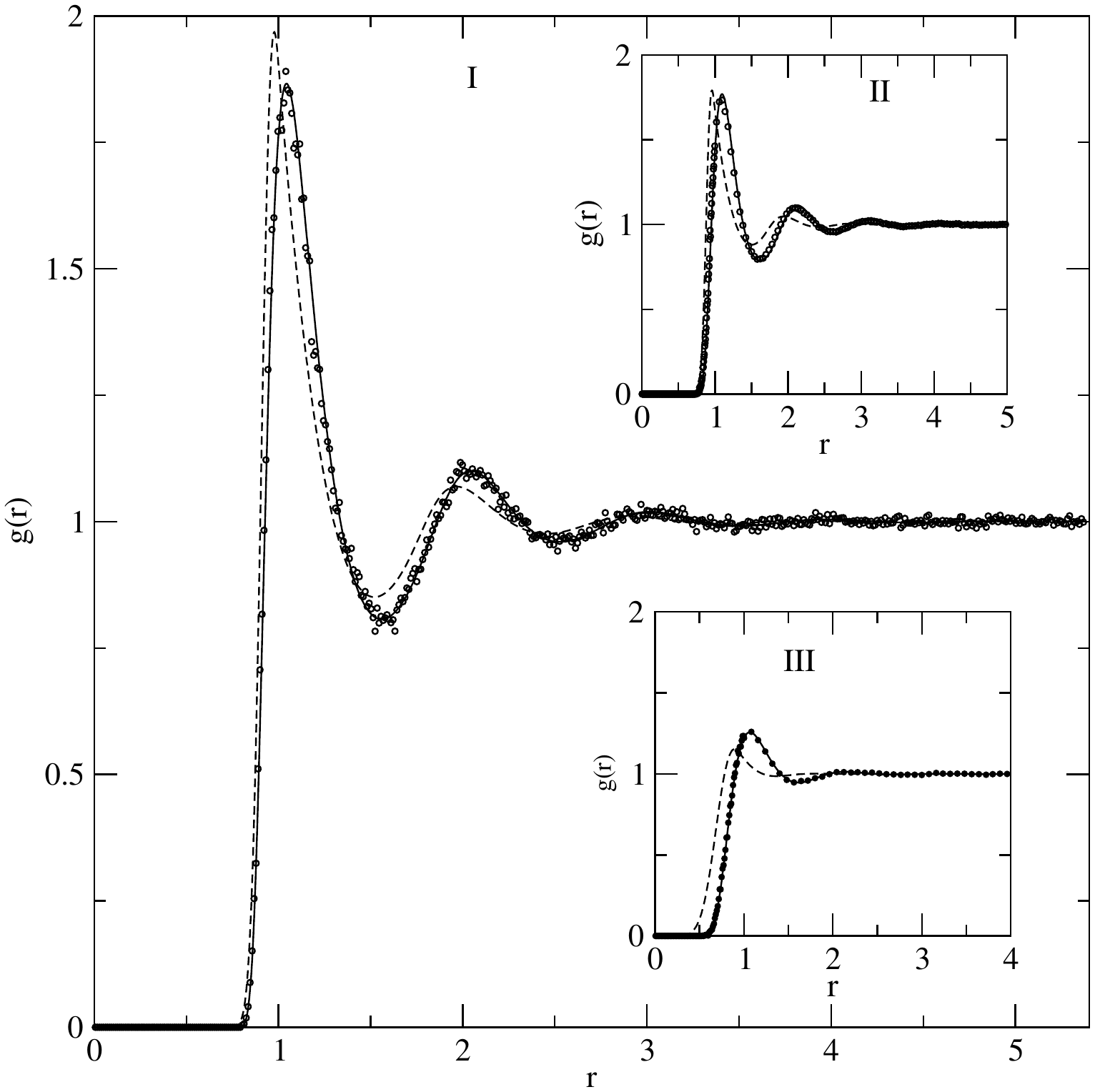}
\caption{Pair correlation function $g(r)$ vs $r/\sigma$, for the
three soft colloids I, II, and III (main panel and the two insets)
interacting through a) soft
sphere (solid) potential defined in equation (\ref{soft-int}) and b)
Hertzian interaction (dashed) potential defined in equation
(\ref{hertz-int}). Parameters for the respective interaction
potentials are given in table~\ref{table04} and table~\ref{table05} for the three soft colloids. The circles in each
figure represent $g(r/\sigma)$ calculated from molecular dynamics
simulation of soft spheres with interaction parameters
$\{n,\epsilon^*\}$ as $\{12.1, 0.571\}$, $\{5.5, 2.340\}$ and
$\{3.1,1.440\}$, respectively, for soft colloids I, II, and III. In
all three cases the relative concentration is
$\zeta/\zeta^*=0.837$.}
\label{fig02}
\end{figure}

\begin{table}[!b]
\caption{Relaxation time data $\tau$ from figure~2b of 
\cite{weitz} in log scale vs the scaled concentration
$\bar{\zeta}=\zeta/\zeta_\mathrm{g}$
            (see text)  for the three soft colloids SC-I, SC-II, and SC-III.}
            \vspace*{5mm}
    \begin{center}
        \vline
        \begin{tabular}{|c|c|c|c|c|c|}
            \hline
            \multicolumn{2}{c|}{~~~~{\bf SC- I}~~~~} &
            \multicolumn{2}{c|}{~~~~{\bf SC- II}~~~~} & \multicolumn{2}{c|}{~~~~{\bf SC- III}~~~~}   \\
            \hline
            ~~~~$\bar{\zeta}$~~~~ & ~~~~$\tau$ ~~~~& ~~~~$\bar{\zeta}$~~~~ &
            ~~~~$\tau$~~~~ &~~~~ $\bar{\zeta}$~~~~ & ~~~~$\tau$ ~~~~ \\
            \hline
            0.016 & -2.972 & 0.016  &   -2.950 & 0.016    &    -2.892 \\
            \hline
            0.188 & -2.872 & 0.154 & -2.596 & 0.101 & -2.462 \\
            \hline
            0.449 & -2.570 & 0.404 & -1.899 & 0.207 & -1.952\\
            \hline
            0.633 & -2.170 & 0.509 & -1.461 & 0.304 & -1.406  \\
            \hline
            0.798 & -1.507 & 0.816 &  0.273 & 0.460 & -0.610 \\
            \hline
            0.904 & -0.447 & 0.890 & 0.904 &0.588 & -0.029\\
            \hline
            0.975 & 1.223 & 0.960 & 1.547 &  0.892 & 1.442\\
            \hline
            1.000 & 2.000 & 1.000  &   2.000 & 1.000   &    2.000 \\
            \hline
        \end{tabular}
        \label{table01}
    \end{center}
\end{table}

\begin{table}[!t]
\caption{The ratio
$\kappa=\zeta_\mathrm{K}/\zeta_\mathrm{g}$, obtained from the
relaxation curves and the corresponding fragility $m$ for the
three respective soft-colloids I, II, and III studied in 
\cite{weitz}.}
\vspace*{5mm}
    \begin{center}
        \begin{tabular}{|c|c|c|c|}\hline
            ~~~~Soft-Colloid~~~~&~~~~The ratio  $\kappa=\zeta_\mathrm{K}/\zeta_\mathrm{g}$~~~~&~~~~Fragility $m$~~~~\\
            \hline
            {\bf I}&1.11  &52.0  \\
            \hline
            {\bf II}&1.70  &12.1  \\
            \hline
           {\bf III}&18.76 &5.3 \\
            \hline
        \end{tabular}
         \label{table02}
    \end{center}
\end{table}

\begin{table}[!t]
\caption{The ratio of the characteristic
            temperatures $T_\mathrm{g}$ and $T_\mathrm{K}$, fragility $m$ of
            the three standard molecular glass forming materials,  are taken
            from~\cite{bohmer1}. The trend is similar to that between the
            ratio $\kappa$ and fragility $m$ as shown in the last two columns
            of table~\ref{table02} for the soft colloids.}
            \vspace*{5mm}
    \begin{center}
        \begin{tabular}{|c|c|c|c|}\hline
            ~~~~Molecular Glass~~~~&~~~~The ratio $T_\mathrm{g}/T_\mathrm{K}$~~~~&~~~~Fragility $m$~~~~\\
            \hline
            Salol& 1.13 & 66 \\
            \hline
            Glycerol& 1.34 & 48 \\
            \hline
            Silica& 3.26& 25 \\
            \hline
        \end{tabular}
  \label{table03}
    \end{center}
\end{table}

The relaxation times data for the three samples as obtained
from figure 2b of~\cite{weitz} are shown in table~\ref{table01}. The range of concentrations $\zeta$ over which the
relaxation time data are provided for the three respective
soft-colloid systems-I, II, and III, are distinct from each
other.  Relaxation behavior for each of the three materials is
shown in figure~\ref{fig01} with respect to relative concentration.
The three curves by construction meet at the same point on the y
axis at $x=1$.  It should also be noted that the relaxation curves
in the Angell plot do not exactly meet at one  point in the low-density side, though it might visually appear to be so  in figure
\ref{fig01}.  On a linear scale, it appears to be at $x=0$, though
in reality these represent very small (nonzero) values of the
packing fraction. The set of points  representing $\tau_\alpha(x)$
vs $x=\zeta/\zeta_\mathrm{g}$ for the respective soft colloids are
fitted the Vogel-Fulcher formula in the form (\ref{rel-fit}).
Each of the relaxation curves corresponding to the three materials
I, II, and III is characterized by a corresponding set of values
for $\{A,\kappa\}$ with the value for parameter ${\cal B}$ kept
fixed at 5. We find that these two parameters follow a linear fit
$A=4.92\kappa-4.88$. The corresponding fragility indices $m$ were obtained from the slope of these curves at the end point,
 i.e., $\zeta=\zeta_\mathrm{g}$. The ratio
$\kappa=\zeta_\mathrm{K}/\zeta_\mathrm{g}$ for each curve and the
corresponding fragility for the three respective soft colloids
obtained this way are listed in table \ref{table02}.
The present scheme of determining fragility of the soft colloids
only works for those materials corresponding to which the
relaxations curves follow the form seen in an Angell plot, with a
chosen value of the parameter ${\cal B}$ (which  defines the so-called glass transition point $\zeta_\mathrm{g}$).  Thus, if the
relaxation does not follow this form for certain materials, such
definitions for fragility do not work.
To compare the relative trends in these  results  with respect
to standard molecular glasses, in table \ref{table03} we show  the
fragility $m$ and the ratio $T_\mathrm{g}/T_0$ for three standard
glass-forming materials, Salol, Glycerol and Silica \cite{bohmer1}
respectively. Here $T_0$ is the temperature in the Vogel-Fulcher
formula $\tau_\alpha=\tau_0\exp [A/(T-T_0)]$ signifying the
divergence of relaxation time and it is determined for the three
materials listed above from the relaxation data presented in~\cite{sokolov-nature}. The temperature $T_\mathrm{0}$ is
considered  similar to the Kauzmann point $T_\mathrm{K}$ for a
glass forming material.
Comparing the results listed in table \ref{table02} with that of
table \ref{table03} we note that with respect to fragility $m$,
variation of $\zeta_\mathrm{K}/\zeta_\mathrm{g}$ in a colloidal
glass is similar to that for $T_\mathrm{g}/T_\mathrm{K}$ in
molecular glass formers. Hence, the trends seen in standard glass
forming materials is the same as that for soft colloids.

\section{High-frequency elastic constants}

In this section, we consider the theoretical models for  high-frequency or short time elastic constants for the soft colloids
and the computation of these properties in terms of the pair
correlation function.  As a test, the input pair correlation
function $g(r)$  is obtained in the present work with the
choice of two different interaction potentials for the soft
colloids.

\subsection{The model}

The Mountain-Zwanzig \cite{mz,mz2,schf66} theory calculates these
quantities  in terms of the interaction potential. Instantaneous
response to stress in the disordered state is related to the
corresponding frequency dependent viscosities \cite{boon}. Thus,
the shear and bulk modulus are obtained from the generalized
viscosities in the high-frequency limit and is denoted as
$G_\infty(\zeta)$ and $K_\infty(\zeta)$ in what follows. The high-frequency limits of the respective elastic constants are obtained
as
\bea
\label{zw-Gw}
G_\infty &=& \ri \lim_{\omega\rightarrow\infty} \{\omega\eta(\omega) \}, \\
\label{zw-Kw} K_\infty &=&
K_0+\ri\lim_{\omega\rightarrow\infty}\{ \omega\eta_V(\omega) \} ,
\eea
where $\eta(\omega)$ and $\eta_V(\omega)$ are respectively the
frequency-dependent shear and bulk viscosities for the fluid.
$K_0$ is the zero frequency bulk modulus obtained in terms of the
adiabatic derivative
\be \label{K0} K_0=-V {\left (
\frac{\partial{P}}{\partial{V}}\right )}_s.
\ee
For pairwise
additive form of the interaction potentials, the total potential energy
$U$ is expressed as a sum of two body terms and the averages on the
right hand side of equations (\ref{zw-Gw}) and (\ref{zw-Kw}) are
expressed in terms of pair correlation function
$g(r)$~\cite{hansen}.
\bea \label{ZM-Ginf} G_\infty &=& \rho_0
k_\text{B}T+\frac{2\piup}{15}\rho_0^2
\int_0^\infty \rd r  g(r) \frac{\rd}{\rd r} \left [ r^4 \frac{\rd u}{\rd r}
\right ], \\
\label{ZM-Kinf} K_\infty &=& \frac{2}{3} \rho_0
k_\text{B}T+P+\frac{2\piup}{9}\rho_0^2 \int_0^\infty \rd r  g(r) r^3
\frac{\rd}{\rd r} \left [ r \frac{\rd u}{\rd r} \right ], \eea
where $\rho_0$ is the density and $P$ is the thermodynamic
pressure for the simple model considered here and is identified
with osmotic pressure for colloids. The latter is obtained with
the standard expression~\cite{hansen}
\be \label{virser} P=\rho_0 k_\text{B}T - \frac{2\piup}{3}\rho_0^2
\int_0^\infty \rd r  g(r) r^3 \frac{\rd u}{\rd r}. \ee
The Poisson's ratio has often been
linked to relaxation behavior of a glass-forming system
\cite{sokolov-nature,oster,sun,ngai,nemilov,johari}.
The Poisson ratio $\nu$  for the soft colloid material is defined in terms of the
ratio of the elastic constants $\mu=K_\infty/G_\infty$ as,
\be \label{PRatio} \nu = \frac{3\mu-2}{2(3\mu +1)}. \ee

In this paper, we consider the elastic response of the soft
colloids over short times or equivalently, in the high-frequency
limit.  For this analysis, we use the data for three soft
colloids  in terms of the shear modulus $G_\mathrm{p}(\zeta)$
evaluated at frequency $\omega_p$. This frequency signifies
the time scale of local particle motion. We identify
the latter  with short time or high-frequency response of the
soft-colloid material and compare it with the corresponding high-frequency
elastic constants obtained from the theoretical models  based on
microscopic statistical mechanics. The observed behaviours of the
data of the elastic constant $G_p$ for the three respective
soft-colloids are also reported~\cite{weitz} to be qualitatively
similar to that of the high-frequency shear modulus $G_\infty$ for
a set of glass forming (molecular) liquids. Hence, we use the
simple model outlined above for the high-frequency elastic
constant of a liquid in terms of the pair correlation function
$g(r)$. We calculate both high-frequency shear modulus $G_\infty$
and bulk modulus $K_\infty$, respectively,  for the soft colloids.
The  short-time elastic response of the liquid is described above
with equations (\ref{ZM-Ginf})--(\ref{PRatio}) in which the
corresponding pair correlation functions are obtained using the
microscopic two-body interaction potential as input. This also makes it possible to
obtain the Poisson's ratio $\nu$ for the respective materials.
Using the expressions on the right hand side of equations
(\ref{ZM-Ginf})--(\ref{virser}) we obtain the two elastic constants
as follows:
\bea \label{Greln} G_\infty  &=&1 +  {\cal I}_1+{\cal I}_2, \\
\label{Kreln} \frac{3}{5} K_\infty &=& 1 + {\cal I}_1-{\cal I}_2,
\eea
where the integrals ${\cal I}_1$ and ${\cal I}_2$
are obtained as functionals of the pair
correlation function $g(s)$ and the dimensionless form of the
interaction potential $\beta{u}(r)\equiv\bar{u}(s)$ as
follows:
\bea \label{int1} {\cal I}_1 [g,\bar{u}] &=&
\frac{4}{5}\zeta \int_{0}^{\infty} \rd s
g(s) s^3\frac{\rd}{\rd s}\left[s\frac{\rd\bar{u}(s)}{\rd s}\right ], \\
\label{int2} {\cal I}_2 [g,\bar{u}] &=&
\frac{12}{5}\zeta \int_{0}^{\infty} \rd s
{g(s)s^3}\frac{\rd\bar{u}}{\rd s}.
\eea
In the above definitions, the radial distribution
function $g(s)$ is expressed in terms of radial distance
$r$ being scaled with a characteristic scale of length $\sigma$
for the chosen interaction potential.
The energy $u$ is scaled with $k_\text{B}T=\beta^{-1}$.
Note that the data reported in \cite{weitz} are only useful
for comparison of the shear modulus $G_\infty(\zeta)$. As
indicated above we consider here two possible bare interaction
potentials to model the soft colloid materials. The characteristic
parameters for the corresponding interaction potential are
adjusted to fit the shear modulus data.  The two interaction
potentials and calculation of the corresponding pair correlation
functions in the respective cases are briefly sketched below.

\subsection{The interaction potentials}

The two-body interaction potentials respectively used to model the
soft colloids are as follows.

\noindent 1. {\em The Hertzian potential}: Interaction between the
colloid particles is given by
\be \label{hertz-int} u(r)=\epsilon_0 {\left
    (1-\frac{r}{r_\mathrm{Hz}}\right )}^{5/2}~~,
\ee
in terms of two positive parameters
$\{\epsilon_0,r_\mathrm{Hz}\}$. The length $r_\mathrm{Hz}$
represents a characteristic length scale at which the interaction
potential $u(r)$ becomes zero and is used as an adjustable
parameter for the potential. The potential energy $u(r)$ is
expressed in units of $k_\text{B}T$ as $\bar{u}(s)$ in terms of the
scaled variable $s=r/\sigma$, where $\sigma$ is the length scale
with which we define the concentration variable as
$\zeta=\piup\sigma^3/6$.
\be \label{hertz-scin} \bar{u}(s)= {\epsilon_0^*}{\left
    (1-\frac{s}{s_\mathrm{Hz}}\right )}^{5/2}~~, \ee
where $s_\mathrm{Hz}=r_\mathrm{Hz}/\sigma$. The energy scale
$\epsilon_0$ in units of $k_\mathrm{B}T$, is obtained as
$\epsilon_0^*= \epsilon_0/(k_\mathrm{B}T)$.\\

\noindent 2. {\em The soft sphere interaction. }
The potential \cite{km1,km2} is given by,
\be
\label{soft-int} u(r)=\epsilon_0 {\left
    (\frac{\sigma}{r}\right )}^n,
\ee
in terms of two positive parameters $\{\epsilon_0,n\}$.
$\sigma$ represents a characteristic microscopic length for the
interaction potential. The soft sphere potential defined in equation
(\ref{soft-int}) is written in units of $k_\text{B}T$ using the scaled
variable $s=r/\sigma$ as,
\be
\label{soft-scin}
\bar{u}(s)= \frac{\epsilon_0^*}{s^n}\,, \ee
where  $\epsilon_0^*=\epsilon_0/(k_\mathrm{B}T)$. In the
$n{\rightarrow}\infty$ limit, the soft sphere potential reduces to
the hard-sphere limit with the hard-sphere diameter
$d{\equiv}\sigma$, and the concentration variable
$\zeta=\piup{\rho}\sigma^3/6$ becomes identical to the packing fraction
$\phi$. For finite $n$, we have soft spheres which are identified with
the deformable colloids.

\begin{figure}[!t]
\centering
\includegraphics[width=0.47\textwidth]{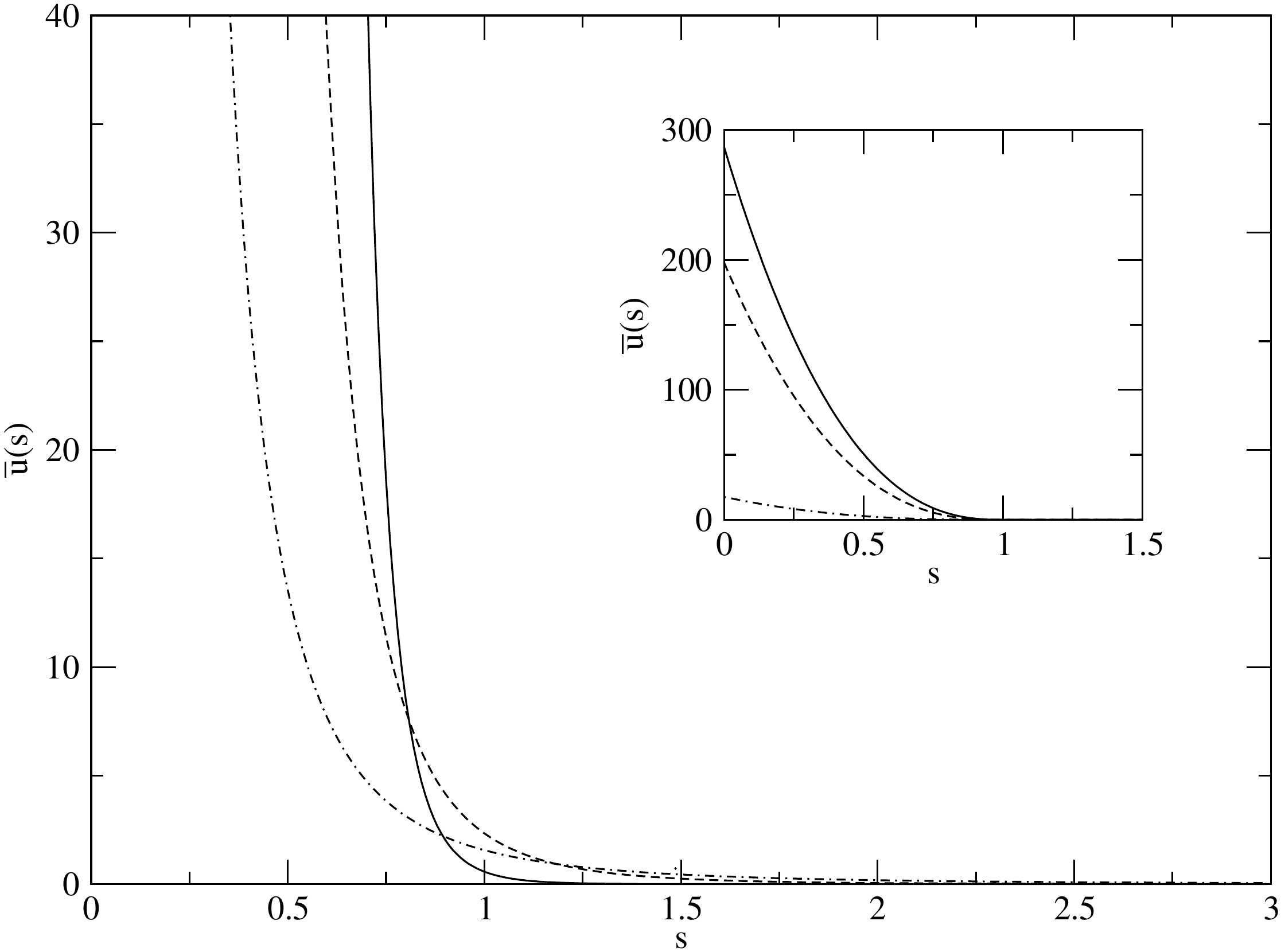}
\caption{The soft sphere potential $\bar{u}(s)$ vs $s=r/\sigma$
with the characteristic parameters for interaction listed in table
\ref{table04} for soft colloids  I (solid), II (dashed), and III
(dot-dashed), respectively. The inset shows the plot of the
corresponding $\bar{u}(s)$ vs $s$ for the Hertzian potential
having characteristic parameters listed in table
\ref{table05}.}\label{fig03}
\end{figure}

The well-known Bridge function method is used to compute the pair
correlation $g(s)$ for each of the above two types of interaction
potentials. These  are  briefly described in Appendix \ref{app1}.
In the soft sphere potential case, the pair function is calculated
using two equivalent approaches. First, we show the theoretical
results obtained from the bridge-function method of Rodger and
Young \cite{rodger}. Second, the static correlations are obtained
through a direct computer simulation of the corresponding fluid.  As
an example, in the main panel of figure \ref{fig02} we show the
respective $g(r)$ vs  $r/\sigma$  for all the three materials
studied at $\zeta/\zeta^*=0.837$. The main
panel shows the results for soft-colloid material I, while the two insets
 show the same respectively for soft colloids II and III. In
each case of the main panel and the insets of  figure \ref{fig02},
we show the $g(r)$  both for the Hertzian potential and for the
soft spheres at the same value of $\zeta/\zeta^*$. The plot of
$\bar{u}(s)$ vs $s$ for the soft sphere potentials corresponding
to the three materials I, II, and III is displayed in the main
panel figure \ref{fig03}. The same results for the Hertzian
potential are shown in the inset of figure \ref{fig03}. In table
\ref{table05},  for each of the respective soft colloid
materials I, II, and III, we compare the lengths $s_\mathrm{Hz}$ for Hertzian
potential and the sizes ($R_0$) of the corresponding spheres\cite{weitz}.

\section{Fragility vs elasticity}

The relaxation and elastic properties for soft colloids are
analyzed to determine their correlations as supported from
experimental data. The high-frequency shear moduli $G_\infty$ for
three soft colloid materials are fitted with the formulae
presented in equations (\ref{Greln})--(\ref{int2}). The pair
correlation function $g(s)$ is the key input in this and is
determined by the interaction potential between the colloidal
particles.
Note that for  the soft sphere potential in the form,
$\bar{u}(s)\sim s^{-n}$, the two integrals ${\cal I}_1$ and ${\cal
I}_2$, respectively obtained in (\ref{int1}) and (\ref{int2}),
are related simply as ${\cal I}_1=-(n/3){\cal I}_2$. In this
case, the elastic constants $G_\infty(\zeta)$ and $K_\infty(\zeta)$
given in equations (\ref{ZM-Ginf})--(\ref{ZM-Kinf}) respectively reduce
to the forms $ G_\infty = 1 + \Delta(n-3)/5$, and $K_\infty=5/3 +
\Delta(n+3)/3$,
in terms of $\Delta=P-1$.
For  $n~{\geqslant}~3$, $G_\infty$  decreases with an increase of
$P$, and is requred for stability.
\begin{table}[!t]
\caption{The list of parameters for the soft-sphere potential defined
in equation (\ref{soft-int}) needed for fitting the data of three soft
colloids I, II, and III. The parameter $\zeta^*$ (column 3) is
obtained through fitting  the elasticity data as shown in
figure~\ref{fig04}--figure~\ref{fig06} (see text). The parameters for the
corresponding soft sphere potential: energy
$\epsilon_0^*=\epsilon_0/k_\text{B}T$ (column 4); softness index
$n$(column 5); $\kappa$ (column 6) is the ratio of the packing at
the glass transition point ($\zeta_\mathrm{g}$) to $\zeta^*$ (see
text); $m$ (column 7) is the fragility index.}
\vspace*{3mm}
    \begin{center}
        \begin{tabular}{|c|c|c|c|c|c|c|c|}\hline
            Soft Colloid&Data Range of&Parameter&energy&softness&
            $\zeta_\mathrm{g}/\zeta^*=$&Fragility \\
            material & Concentration $\zeta$ & $\zeta^*$  &
            $\epsilon^*$ & $n$ & $\kappa$&
             $m$\\
            \hline
            I &0.300-0.540 & 0.500& 0.571 &12.1 & 1.11  & 52.0 \\
            \hline
            II &0.264-0.400 & 0.100 & 2.340 &5.5 & 6.65 & 12.1\\
            \hline
            III &0.270-0.430 & 0.054  & 1.440 &3.1 & 12.46  & 5.3\\
            \hline
        \end{tabular}
 \label{table04}
    \end{center}
\end{table}

\begin{table}[!t]
\caption{The list of parameters for Hertzian potential defined in
equations (\ref{hertz-int})--(\ref{hertz-scin}) needed for fitting the
data of three soft colloids I, II, and III. $\zeta^*$ (column 3)
is obtained through fitting  the elasticity data as shown in
figure~\ref{fig04}--\ref{fig06} (see text). The parameters for
corresponding Hertzian potential : energy
$\epsilon_0^*=\epsilon_0/k_\text{B}T$ (column 4);  $s_\mathrm{Hz}$
(column 5) is the length at which the potential $\bar{u}(s)$ is
zero. The length $R_0$  (column 6) denotes the radius of the
deformable spheres for the three respective soft colloids as
reported in~\cite{weitz}. The lengths $s_\mathrm{Hz}$, and
$R_0$,  are scaled so that for material I, these quantities are
unity by choice. Finally, $m$ (column 7) is the fragility of the
soft colloids shown here for comparison.}
 \vspace*{3mm}
\begin{center}
\begin{tabular}{|c|c|c|c|c|c|c|}\hline
            Soft Colloid&Data Range of&Parameter&energy&range
            &Radius&Fragility  \\
            studied& Concentration $\zeta$ & $\zeta^*$ &
           $\epsilon_0^*$ & $s_\mathrm{Hz}$ 
            & $R_0$(A)&$m$\label{key}\\
            \hline
            I &0.300--0.540 &0.50 & 287 &1.00& 1.00&52.0\\
            \hline
            II &0.264--0.400 &0.10 & 198 &0.98& 0.97&12.1\\
            \hline
            III &0.270--0.430 &0.054 & 17.6 &0.96 &0.84&5.3\\
            \hline
        \end{tabular}       
 \label{table05}
\end{center}
\end{table}

\begin{figure}[!t]
    \centering
    \includegraphics[width=0.4\textwidth]{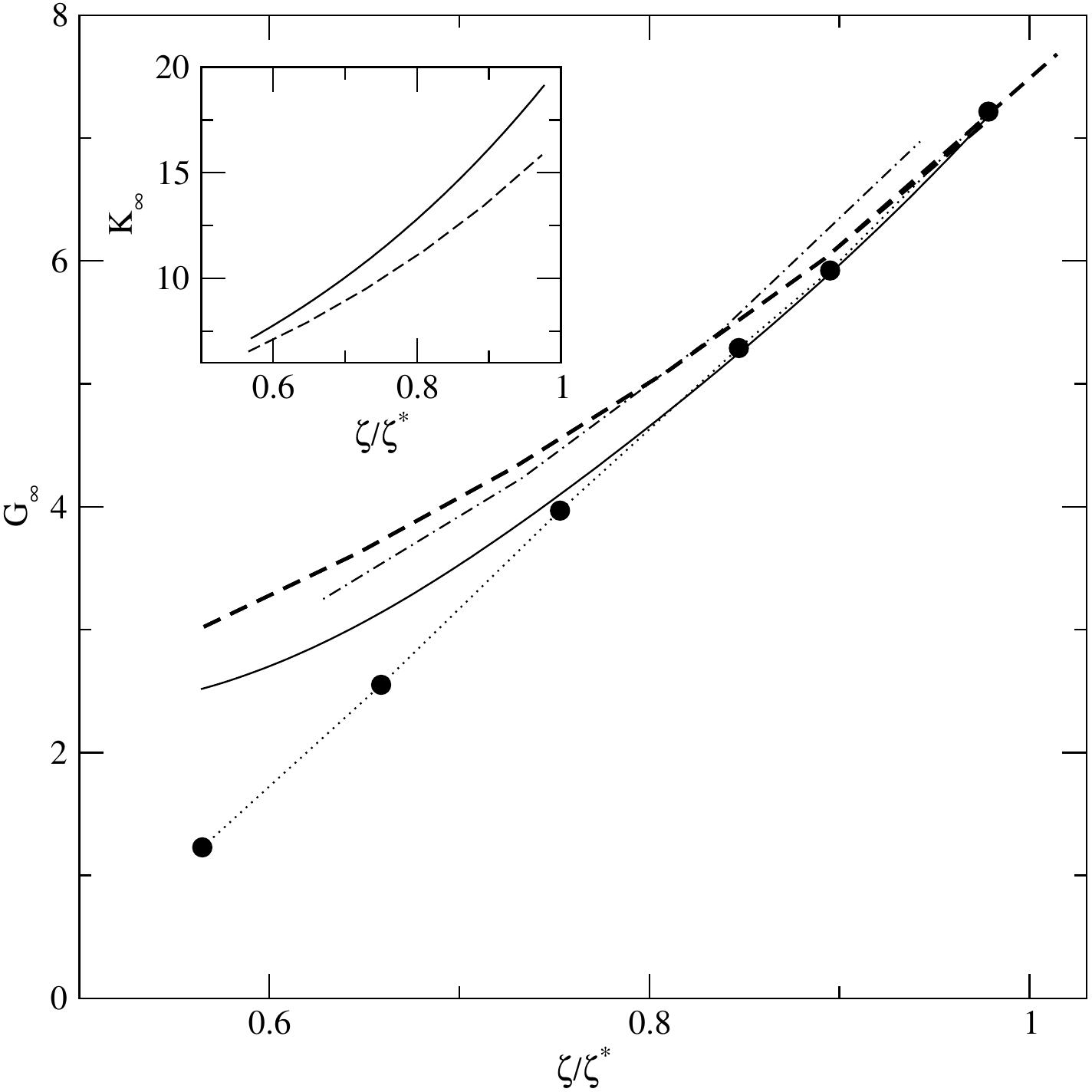}
    \caption{Main panel: Theoretical results for
        shear modulus $G_\infty(\zeta)$ in units of
        $\rho_0{k}_\mathrm{B}T$ vs scaled concentration $\zeta/\zeta^*$
        (see text). We use equation (\ref{Greln}) with structural inputs of
        $g(r)$ obtained from (i) the bridge function method for soft sphere
        interactions(solid line); (ii) molecular dynamics simulations with
        soft sphere interactions (dot-dashed); and (iii) the bridge function
        method for Hertzian potential (dashed).  Comparison with shear
        modulus data for $G_\mathrm{p}(\zeta)$ for soft colloids I taken
        from figure 3b of~\cite{weitz}; data scaled with
        a material-dependent factor for each system:
        filled circles are connected with the dotted line.
        Insets: theoretical results from equation~(\ref{Kreln})
        for bulk modulus $K_\infty(\zeta)$ in units of
        $\rho_0{k}_\mathrm{B}T$ vs scaled concentration $\zeta/\zeta^*$,
        with the same input $g(r)$  as obtained from fitting the shear modulus
        data for the material shown in the corresponding main panel: for
        soft sphere potential (solid line) and Hertzian potential (dashed
        line) bridge function method.}
        \label{fig04}
\end{figure}

\begin{figure}[!t]
    \centering
    \includegraphics[width=0.4\textwidth]{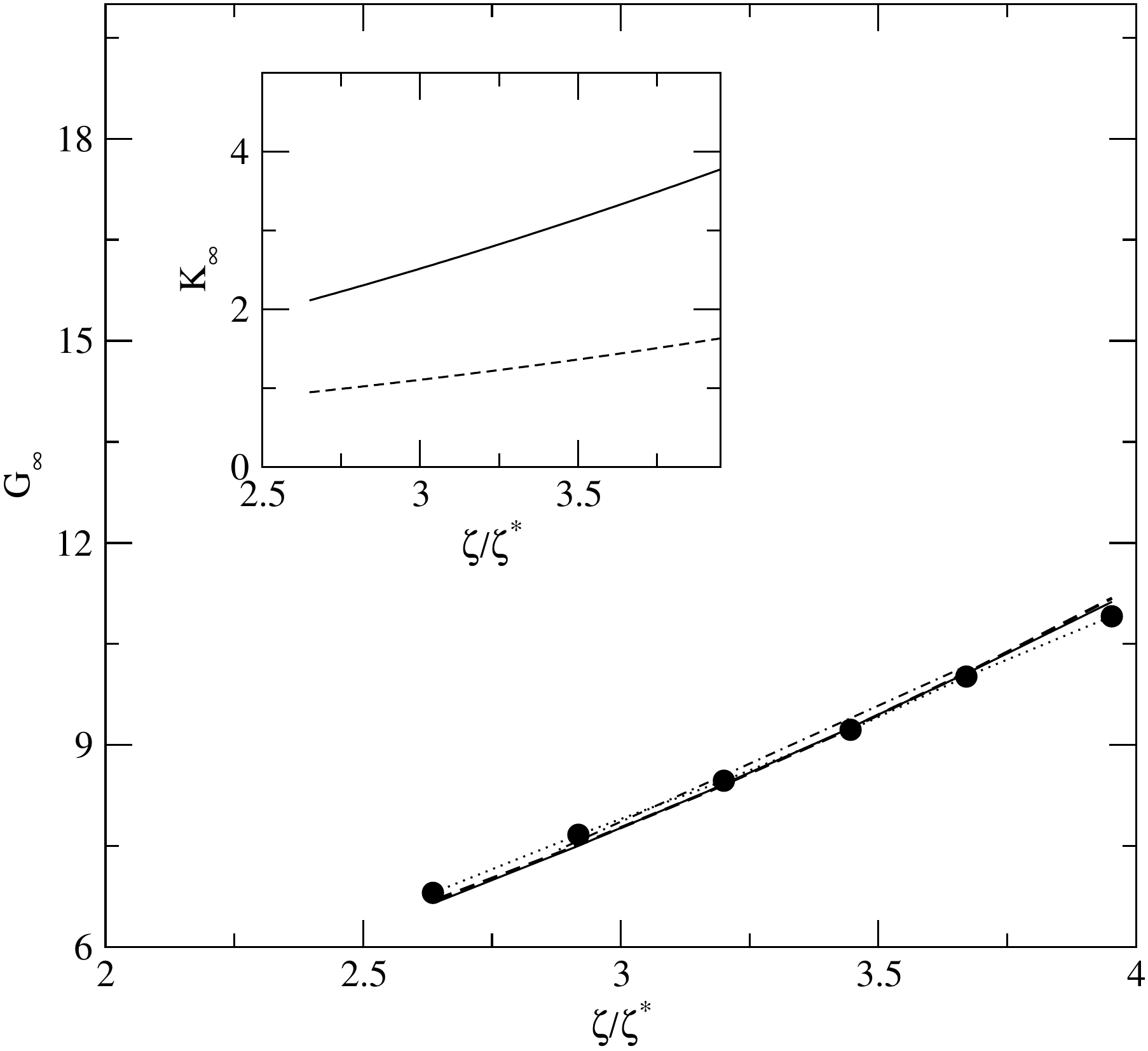}
    \caption{The same as figure \ref{fig04} for soft colloid II of  \cite{weitz}}
    \label{fig05}
\end{figure}

\begin{figure}[!t]
    \centering
    \includegraphics[width=0.4\textwidth]{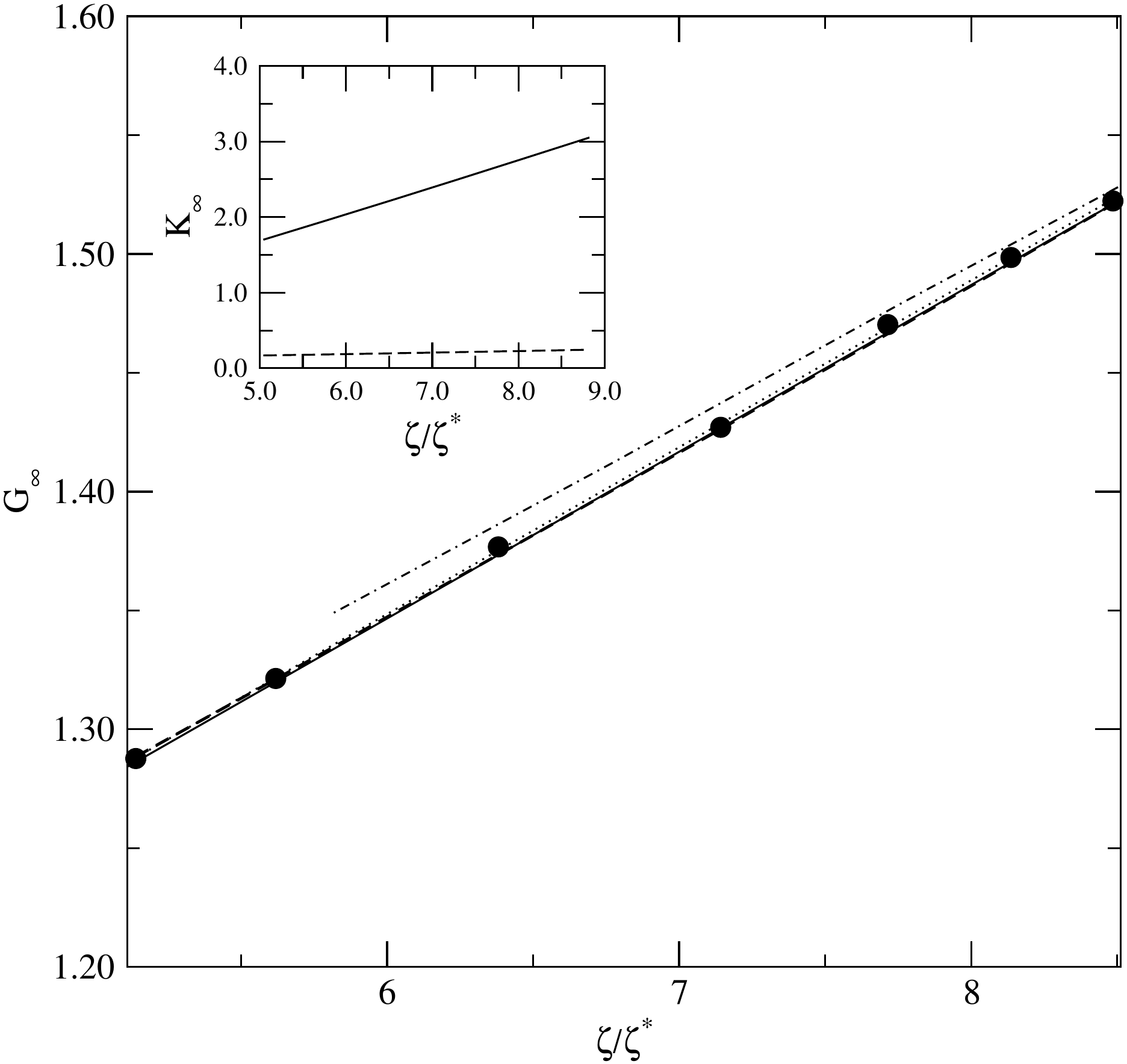}
    \caption{The same as figure~\ref{fig04} for soft colloid III of \cite{weitz}}
    \label{fig06}
\end{figure}

\begin{figure}[!b]
\centering
\includegraphics[width=0.64\textwidth]{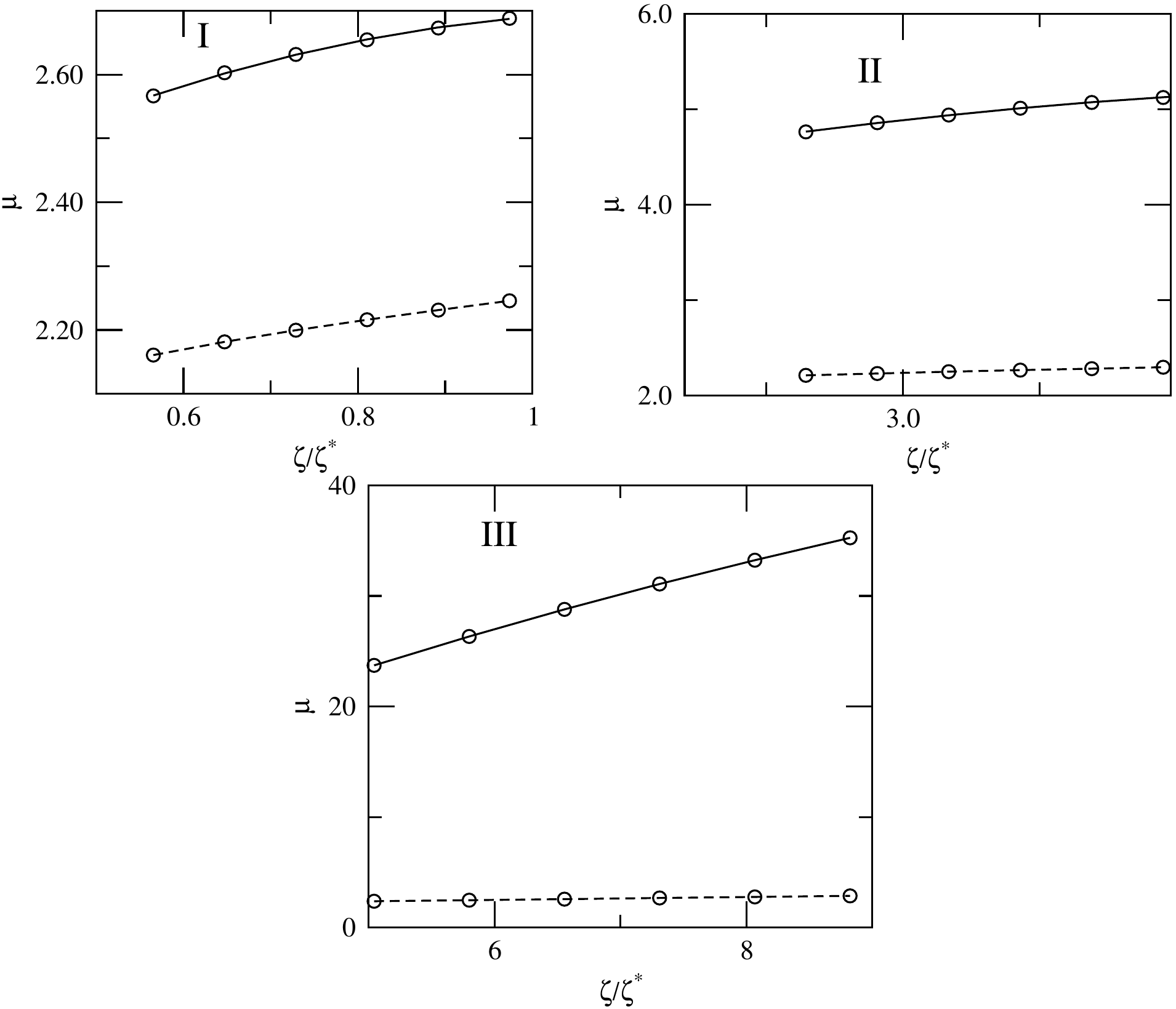}
\caption{The ratio $\mu=K_\infty(\zeta)/G_\infty(\zeta)$ vs
$\zeta/\zeta^*$ for the three materials : I, II, and III. The
results correspond to the calculation done with $g(r)$ for soft sphere
potential (solid) and Hertzian potential (dashed) in each figure.}
\label{fig07}
\end{figure}

For the Hertzian potentials defined in equation (\ref{hertz-scin}),
the shear modulus data are fitted by adjusting parameters,
$\{s_\mathrm{Hz},\epsilon^*_0\}$, while for the soft sphere
potential defined in (\ref{soft-scin}) the corresponding
parameters are $\{ n,\epsilon^*_0\}$. The results are listed in
tables \ref{table04} and \ref{table05}.  For a particular soft
colloid material (I, II, or III), the parameter values
corresponding to a chosen interaction potential (respectively
being Hertzian or soft-sphere) are kept fixed while fitting the shear
modulus data for that material over the whole range of density
values. We obtain the respective set of parameters for the
interaction potential so that the corresponding pair functions
$g(s)$ in the theoretical formulae produce the best fit with the
$G_\infty$ data over the whole density range studied for a
particular soft colloid material. With a suitable choice of the parameter
$\zeta^*$ for each case, we match the $G_\mathrm{p}(\zeta)$ data
for the three materials. Corresponding to the specific set of
parameter values for the potential $\bar{u}(s)$, the pair
functions $g(s)$  (at different densities) fit the shear moduli
$G_\mathrm{p}(\zeta)$ data. In figure~\ref{fig04}--figure~\ref{fig06}, we
show the comparison between experimental data for $G_\infty$ of
the three respective soft colloids with the corresponding
theoretical results obtained using a) the Hertzian and b) the soft
sphere interaction potentials.
Next, the same $g(r)$ in each respective case is used to obtain
the corresponding  bulk modulus $K_\infty(\zeta)$. This procedure
is followed  for each of the three Soft colloids, using the
formulae  (\ref{ZM-Kinf}), (\ref{Kreln}), and
(\ref{int1})--(\ref{int2}). The bulk moduli $K_\infty$ 
obtained for the three samples SC-I, II, and III are shown
respectively in the insets of  figure~\ref{fig04}--figure~\ref{fig06}.
For all three materials I, II, and III, the bulk modulus obtained
for the Hertzian potential case is lower than the corresponding
results for the soft sphere potential. In figure \ref{fig07} we show
how the ratio $\mu=K_\infty(\zeta)/G_\infty(\zeta)$ changes with
concentration $\zeta$ for the three soft materials in I, II, and
III. Since the ratio $\mu$ is material dependent, it is not very
sensitive to an increase of the concentration $\zeta$.

\begin{figure}[!t]
    \centering
    \includegraphics[width=0.47\textwidth]{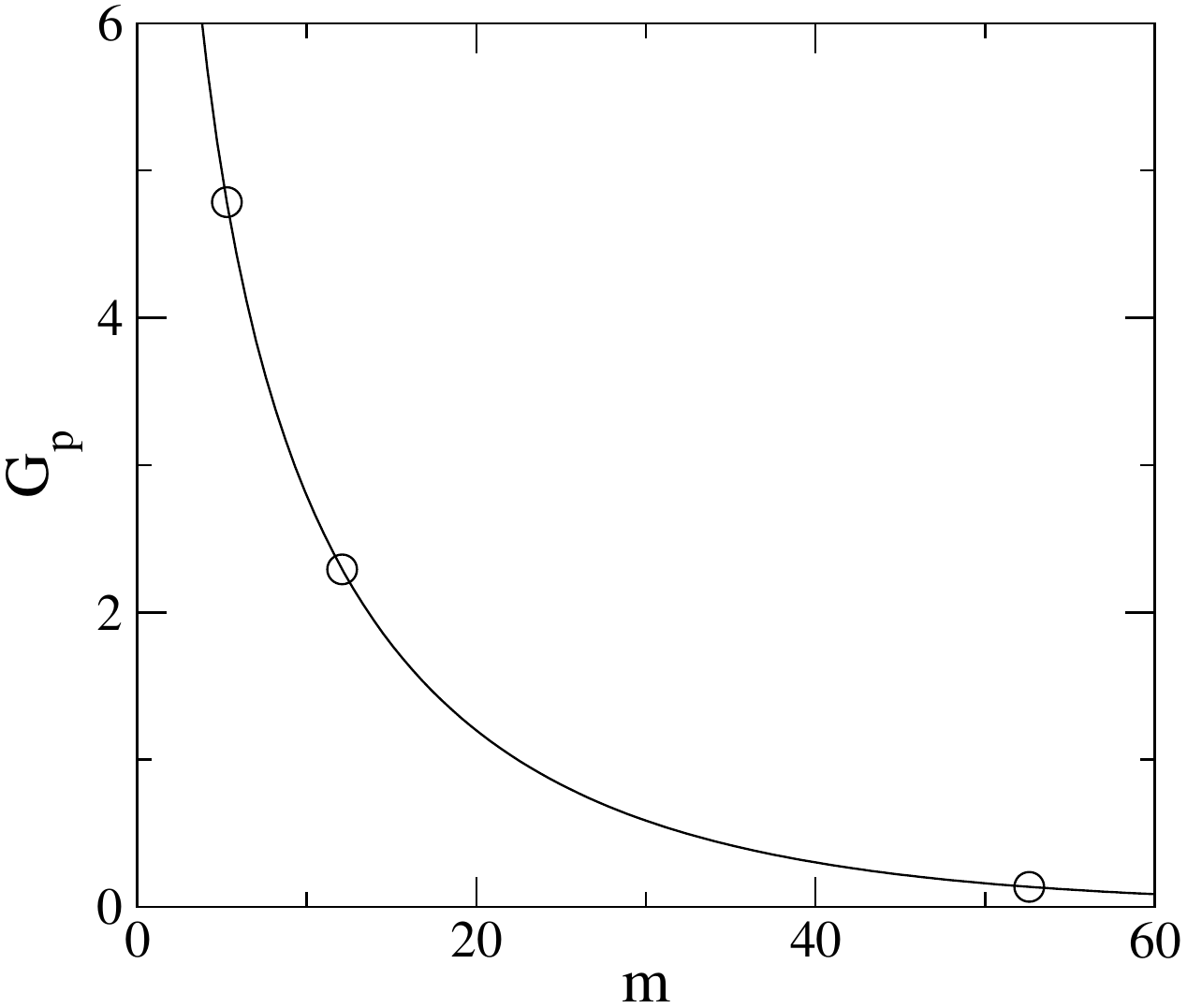}
\caption{The plot of the shear modulus $G_p$ vs fragility $m$ for the
three soft colloids I, II, and III (see text). Data taken from
figure~3a of~\cite{weitz} and scaled with a factor of
${\tilde{\sigma}^3}/(k_\text{B}T)$. Here $\tilde{\sigma}=0.1R_0$ (see
text).}\label{fig08}
\end{figure}

\begin{figure}[!t]
\centering
\includegraphics[width=0.47\textwidth]{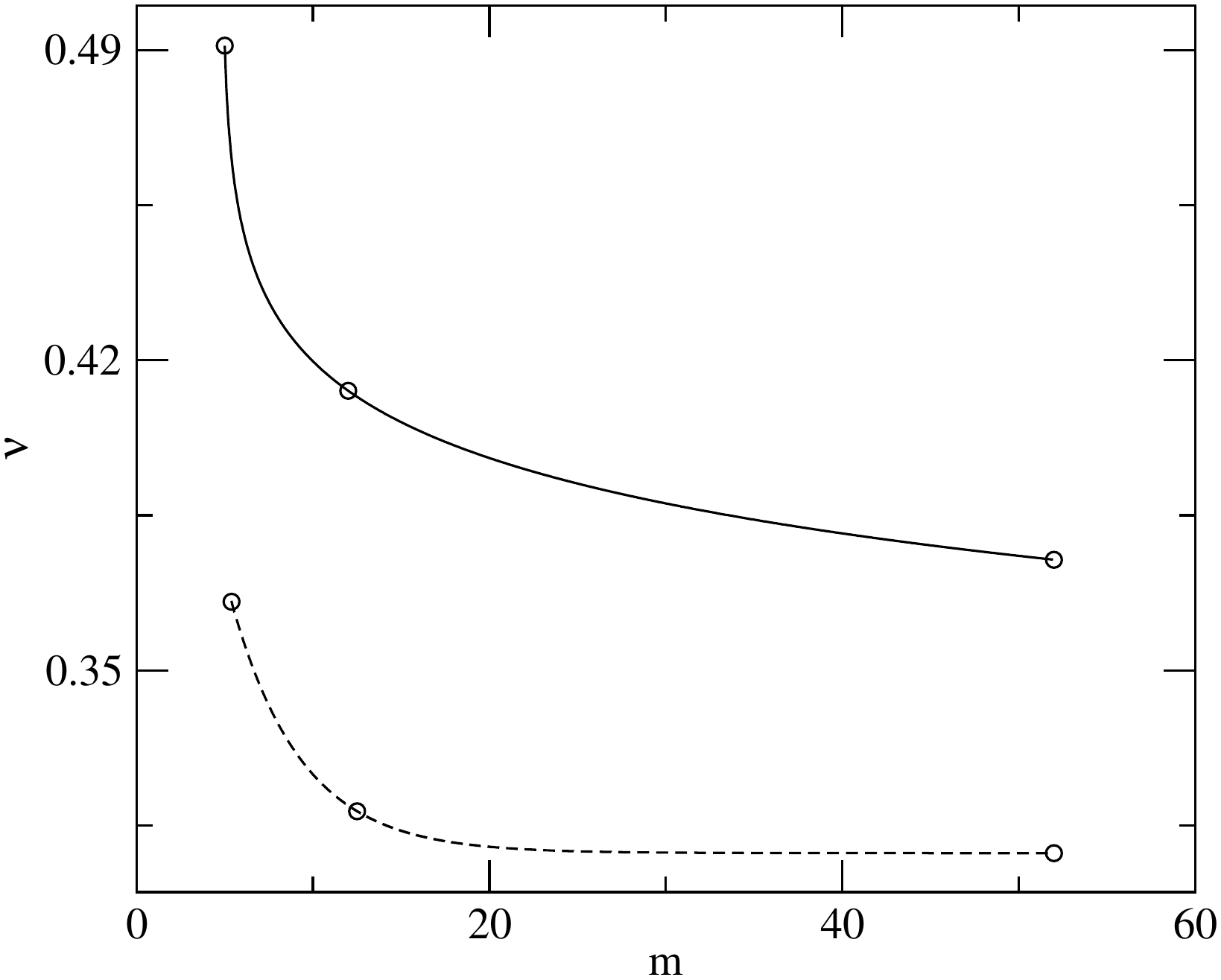}
\caption{The plot of the Poisson's ratio $\nu$ vs the fragility
$m$ for the three materials studied in the present work, for the
soft sphere potential (solid) and Hertzian potential (dashed).}\label{fig09}
\end{figure}

The optimum parameter values for the respective interaction
potentials, obtained by matching shear modulus data with the
theoretical formula (\ref{ZM-Ginf}), signify  a  particle size
for the respective materials and correlate corresponding
radii $R_0$. The  cutoff length $s_\mathrm{Hz}$ of the Hertzian
potential, and $R_0$,  are shown in table~\ref{table05}. The
results shown are scaled so that for material I, these
quantities are unity.  The respective variations of
$s_\mathrm{Hz}$ and $R_0$ with the fragility $m$ of the three soft
colloids I, II, and III show the same qualitative trend.
Similarly,  the softer is the potential, the lower is the value of $m$. With the soft sphere potential, for increasing values of
softness index $n$, the fragility $m$ increases. This also agrees
with the view that hard-sphere systems ($n\rightarrow\infty$) are
most fragile. In figure \ref{fig08} we present how the shear modulus
for the respective three soft colloids I, II, and III correlates
with the corresponding fragility values. This figure presents a
plot of ${G_\infty}$ (expressed in units  $k_\text{B}T/\tilde{\sigma}^3$)
at $\zeta=\zeta_\mathrm{g}$ vs the fragility $m$ for the
corresponding material. The experimental shear modulus data
displayed here are taken from figure~3a of \cite{weitz} and
these form the common basis for finding the optimum parameter
values for both types of potentials. The trend of correlation of
$m$ with $G_\infty$ is shown here is similar to that of the Poisson's
ratio which we  discuss next. We calculate the Poisson's ratio
$\nu$ using the formula equation~(\ref{PRatio}) at the glass transition
point $\zeta=\zeta_\mathrm{g}$. In figure \ref{fig09} a plot of
$\nu$ at the glass transition point $\zeta_\mathrm{g}$ with
respect to the corresponding value of the fragility index $m$ is
shown for both types of interaction potentials.

\begin{figure}[!t]
    \centering
    \includegraphics[width=0.5\textwidth]{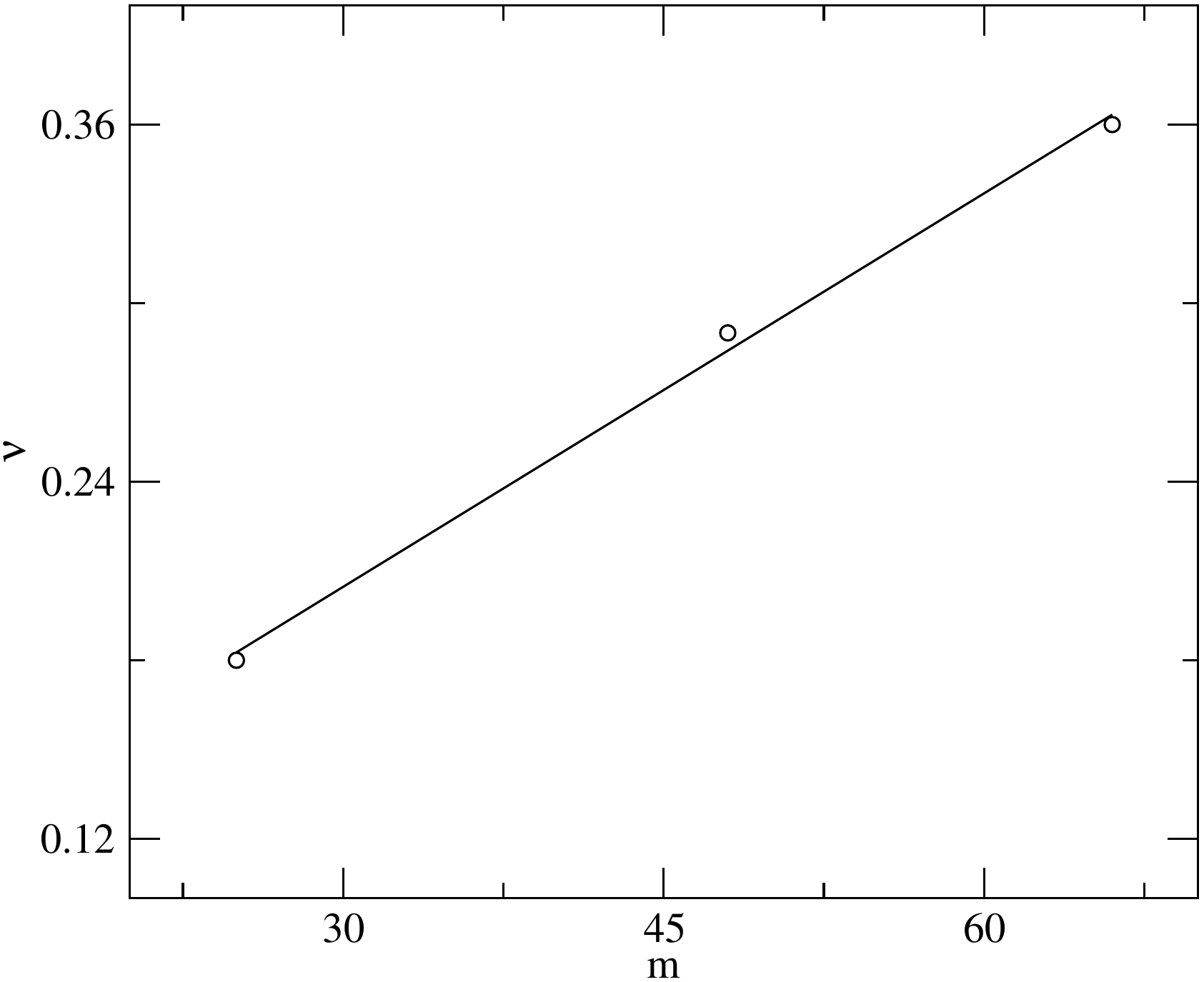}
\caption{The plot of the Poisson's ratio $\nu$ vs the fragility
$m$ for the three molecular glass formers. In the increasing order
of fragility (m)  these  are Silica, Glycerol, and Salol taken
from~\cite{bohmer-angell}. The results in each case are shown
as points and the lines are a guide to the eye. The trends of the
data are opposite for the molecular systems and soft
colloids.}\label{fig10}
\end{figure}

In all cases,  corresponding to both types of interaction
potentials, soft sphere repulsive and Hertzian, the Bulk modulus
is larger than the corresponding shear modulus. However, for the
soft sphere interaction case, the ratio is much larger and hence
the Poisson ratio remains close to $0.4\sim{0.5}$. For the Hertzian
potential, the difference between bulk and shear moduli is less
pronounced and hence the variation in Poisson's ratio is wider.
The qualitative trends for correlations of either $G_\infty$ or
$\nu$ with the fragility is observed to be same. For the sake of
comparison, figure \ref{fig10} shows the correlation between the
same quantities, $\nu$ and $m$ in molecular systems. Here, the
results for three standard molecular glass-forming systems
\cite{weitz} as taken from the literature
\cite{bohmer-angell,sokolov-nature,novikov2} are displayed. In
soft colloids,  Poisson's ratio $\nu$ varies with respect to $m$
in a manner which is {\em opposite} to what is observed in
molecular glass-forming systems. These opposing trends are
discussed further in the next section.

\section{Discussion}

The static and dynamic behaviours of colloidal hard-sphere systems
\cite{pnpusey,lowen,roji} have similarities with  those seen in
molecular liquids.  Experimental studies of Mattsson et al.
\cite{weitz} demonstrated that even deformable colloidal
suspensions exhibit the dynamical behaviour qualitatively similar  to
a fluid of non-deformable spheres. Relaxation times grow with
concentration ($\zeta$) of constituent particles in a soft colloid material
and the concept of fragility is directly extended to suspensions
of deformable  colloidal particles in terms of density.  We
identify each of the  relaxation curves displayed in the so-called
``Angell-plot''  with a corresponding set of parameters for a
(chosen) type of two-body  interaction-potential. We do this by
fitting  the short time (high-frequency)  shear modulus
data~\cite{weitz} for the respective materials using  theoretical
formulae involving pair correlation function $g(r)$ for the
system. The appropriate $g(r)$ for each soft-colloid is chosen by
adjusting the parameters for interaction potential. We have chosen
in our study the Hertzian potential as well as soft sphere
potential.  For the same pair potential, the bulk moduli
$K_\infty(\zeta)$ are calculated in each case and hence the
Poisson's ratio is calculated.

It will be useful to pinpoint the key approximation involved in the
simple model for the soft colloids as we presented here. The
system of deformable soft spheres is modelled here in terms of a
system of microscopic particles interacting through a bare two-body  potential. For the latter, we have made the simplest possible
choices, namely soft sphere and Hertzian types. Thus, the
deformable property of the spheres is mapped in terms of the
softness of respective  interaction potentials, characterized by
the corresponding  strength and range for each system.  The
elastic constants or pressure, respectively given in equations
(\ref{ZM-Ginf})--(\ref{ZM-Kinf}), and (\ref{virser}) are expressed
in terms of the two point pair correlation function $g(r)$ for the 
corresponding interaction potential. With the appropriate choice of interaction
parameters in the respective cases,  obtained by fitting the shear
modulus data, the number density or concentration  $\zeta$ is the key
variable in the theoretical model. In the present work we
approximate the pair correlation function $g(r)$ with a simple
form for an isotropic system with uniform density. Hence, $g(r)$
depends on the parameters of interaction potential and the average
number density. A system of deformable spheres in reality may have
a nonuniform state and the pair correlation function $g({\bf r})$
will have to account for the concentration dependent change and
local packing in a self-consistent manner. This will be a way of
accounting for the variations in the size of particles due to
the local pressure. A self-consistent formulation of pair function in
terms of local density, going beyond  the simple isotropic form
$g(r)$ used here for uniform density state, will be needed to make
a model for the long-time dynamics of the soft colloid materials. In the
present context, we use the simplest model for the short-time
behaviour of the soft colloid materials.

The correlation observed here linking the fragility $m$ and softness
index $n$ or the Poisson's ratio $\nu$ in the glass forming
system, is based on the analysis of  experimental data as well as on the use
of simple models. The theoretical model which links the
interaction potential to elastic constants  in this work, was
originally proposed by Zwanzig and Mountain  to explain the
short time elastic response of the liquid in terms of the
interaction potential.  Deductions of these formulae for the
elastic constants do not require explicit models for the dynamics
for the soft colloids since the theory primarily focuses on the
short time or high-frequency elastic response of the liquid. Time
correlation functions in the Green-Kubo expressions for the
viscosity are only calculated here in the infinite frequency or
zero time limit and hence one is dealing with only static 
correlations. For the time-dependent behaviour or finite frequency
response of soft materials valid over different time scales,
modelling of the dynamics \cite{weitz-prl} would be needed. For
the long time elastic response in a frozen state, mode coupling
contributions \cite{my_book} coming from nonlinear coupling of
density fluctuations \cite{dyre-j} are important. 
For the Brownian dynamics of the soft colloid particles, the relaxation times observed here are long and it indicates that strongly cooperative motions occur.
The glassy dynamics in soft colloids was studied using a
microscopic model, such as the mode coupling theory (MCT)
\cite{erica-ken,my_rmp}. In analogy with these models, we have
chosen soft repulsive spheres in our
theoretical analysis while comparing with experimental data.

For models in which interaction between colloid particles  is
chosen to be soft sphere potential, the relaxation data indicate the
lowering of the fragility index $m$ with decreasing values of
softness index $n$. A similar dependence of the softness of
interaction potential on the fragility of an atomic system was obtained using numerical studies of binary Lennard-Jones
systems \cite{pes-frag}. This trend, however, is not universal.  In
some simulations of binary mixtures \cite{francr}, the fragility
was even reported to be independent of softness of the potential.
Temperature dependence of the diffusion coefficients for various
$n$ collapses in this case onto a universal curve. Thus, as concerns
the link between fragility and softness index, our findings do
match with some, but not all, numerical results obtained in case of
molecular systems. It is also worth to note here that the
fragility for a molecular system is defined with the slope of the
corresponding relaxation curve in the Angell-plot at
$T=T_\mathrm{g}$. In the present work, following~\cite{weitz}, fragility is calculated through density dependence,
at $\zeta=\zeta_\mathrm{g}$. Correlation between the Poisson's
ratio and fragility was found
\cite{oster,sun,ngai,malfait,souri,nemilov,novikov,novikov2,ngai1} to be
non-universal even within only molecular systems. Thus, while there
exist a fundamental link between fragility and elasticity through
the basic interaction potential, and the underlying structure of
the liquid influences its long time dynamics, and the observed
correlation between $m$ and $\nu$ demonstrated in the present work
is still at the level of a hypothesis.

From a theoretical perspective, the observed link between the
fragility of a metastable liquid and its elastic coefficients is a
manifestation of structural effects on the dynamics of a many-particle system. The elastic response of the fluid approaching the glass
transition signifies a solid-like behavior. Fragility index, on the
other hand, is related to the relaxation process in the liquid
state  near the so-called glass transition point. Developing a
common basis for both elasticity and fragility involves
understanding the process of a rigidity-transformation of the
metastable liquid into an amorphous solid-like state
\cite{dyre_rmp}. The success of basic theoretical models in
comprehending this transition is only partial so far.

\section*{Acknowledgement} AM acknowledges UGC-BSR fellowship,
and LP acknowledges CSIR, India for financial support. SPD
acknowledges support under JC Bose fellowship grant.
\newpage

\appendix
\section{The pair-correlation function}
\label{app1}

In this Appendix we very briefly describe  the standard procedure
followed for calculation of the pair correlation functions for the
chosen interaction potentials.

\subsection{The Bridge functions for  pair correlation}

We  use here the bridge function method of standard liquid state
theory to obtain $g(r)$ for a  soft sphere system with the chosen
values of the parameters $n$ and $\epsilon_0^*$ . We also compute
the static functions through direct molecular dynamics simulations
for the chosen soft sphere potential. We briefly outline here the
bridge function method as well as the simulation approach which we
follow to compute the pair function $g(r)$. The total correlation
function for the fluid is defined as, $h(r)=g(r)-1$. The  direct
correlation function denoted as $c(r)$ \cite{hansen} is linked to
$h(r)$ by Ornstein-Zernike (O-Z) relation
\be \label{ozeqn} h(r) = c(r) + \int h(r-r{'}) c(r{'}) \rd r{'}.
\ee
The set of equations for $c(r)$, $h(r)$ are closed by choosing a
closure relation through the introduction of a bridge function
$B(r)$. The closure equation is chosen in the form
\be y(r) = \re^{h(r)-c(r)+B(r)}, \ee
where $y(r)$=$g(r)\exp[u(r)/(k_\text{B}T)]$ and $B(r)$ is the bridge
function introduced in defining the closure. To construct $B(r)$,
various approximations have been devised in the literature. For
soft sphere potentials, a successful approach is the
Rodger-Young bridge function \cite{rodger,jcp08}. We solve the O-Z
equation (\ref{ozeqn}) numerically with the Rodger and Young
closure. The corresponding results for $g(r)$ obtained agree well
with the same quantity calculated using computer simulation. Our
results show that the elastic properties are very sensitive to the
pair function.

In the computer simulations, we study a system of particles
interacting {\em via} inverse power law potential given by equation
(\ref{soft-int}). We set a cut off of the potential at
$r_\text{cut}=2.5\,\sigma$ and study the dynamics of one thousand
particles. Using the molecular dynamics simulation, the equilibrated
samples are generated for different state points. The potential
parameters are adjusted to obtain the $g(r)$ that provides,
through the formula (\ref{Greln}), the best fit to the shear
modulus data reported in  \cite{weitz}.  The equilibrated
samples are initially generated for different state points at
$n=12.1,\,T^*=1.75$; $n=5.5,\,T^*=0.43$; and $n=3.112,\,T^*=0.639$
using Nos\'{e}-Hoover thermostat at constant NVT. By evolving
further, using NVE integrator, for each state point at least 100
independent samples well separated by the structural relaxation
times are obtained. The time step used throughout the simulation
is  $0.001$ in Lennard-Jones time units. In all the three state
points, we find that the systems  do not crystallize as shown by
the pair correlation function. The pair correlation function is
calculated by averaging over independent samples.

\ukrainianpart
\title{Кореляція між кінетичною крихкістю і коефіцієнтом Пуассона  на основі аналізу даних для м'яких колоїдів}

\author{A. Мондал\refaddr{label1},
Л. Премкумар\refaddr{label1, label2},  С.П. Дас\refaddr{label1}
} 

\addresses{\addr{label1} Школа фізичних наук, Университет Джавахарлала Неру, Нью-Делі  110067, Індія
\addr{label2} Фізичний факультет, Національний технологічний інститут  Маніпура, Імфал 795004, Індія}

\makeukrtitle 
\begin{abstract}
Ми розглядаємо зв’язок між крихкістю та еластичністю на основі аналізу
 даних для набору м'яких колоїдних матеріалів
що складається з деформованих сфер, описаних в  [Mattsson { et
al.}, Nature, 2009, {\textbf 462}, 83]. 
	У даній роботі проведено
	кількісний аналіз з допомогою явного
	визначення індексу крихкості $ m $ в термінах
	залежності густини, розширюючи відповідну формулу $ m $ для молекулярної
	системи з точки зору температурної залежності.
	Окрім цього,
	 дані для високочастотного модуля зсуву для
	відповідного м'якого колоїду застосовано до відповідного теоретичного виразу
	для того самого модуля. Цей вираз для пружної константи отримано
	на основі відповідної парної кореляційної функції рідини з однорідною густиною. Парну кореляційну функцію встановлюють шляхом відповідного вибору параметрів для
	потенціалу взаємодії двох тіл для відповідного матеріалу м’якого колоїду. Природа кореляції між крихкістю і
	коефіцієнтом Пуассона, що спостерігається для м'яких колоїдів, якісно відрізняється
 від природи молекулярного шкла. 
	Зв'язок, що спостерігається між крихкістю метастабільної рідини та її
	коефіцієнта пружності, є проявом впливу
	будови рідини на її динаміку.
	Таким чином, у даній роботі проаналізовано дані стосовно м'яких колоїдів. За рахунок введення дефініцій зі статистичної механіки, отримано кореляцію між кінетичною крихкістю та
	коефіцієнтом Пуассона для м'яких матеріалів. 	
	
\keywords кінетична крихкість, пружний відгук, часи релаксації, склування
\end{abstract}

\end{document}